\documentclass[iop]{emulateapj}

\shorttitle{AGN in massive galaxies at $z\sim2$}
\shortauthors{Olsen et al.}

\usepackage{url}
\usepackage{amsmath,amsthm,amssymb}
\usepackage{multirow}

\begin{document}

\title{Evidence for Widespread Active Galactic Nucleus Activity among Massive Quiescent Galaxies at \emph{z} $\sim 2$}

\author{Karen P. Olsen}
\author{Jesper Rasmussen}
\author{Sune Toft}
\author{Andrew W. Zirm}
\affil{Dark Cosmology Centre, Niels Bohr Institute, University of Copenhagen, Juliane Maries Vej 30, DK-2100 Copenhagen, Denmark; karen@dark-cosmology.dk}

\begin{abstract}
We quantify the presence of active galactic nuclei (AGNs) in a mass-complete ($M_{\ast}>5\times10^{10}M_\odot$) sample of $123$ star-forming and quiescent galaxies at $1.5\leq z\leq 2.5$, using X-ray data from the 4 Ms Chandra Deep Field-South (CDF-S) survey. 
$41\%\pm7\%$ of the galaxies are detected directly in X-rays, $22\%\pm5\%$ with rest-frame 0.5-8\,keV luminosities consistent with hosting luminous AGNs ($L_{0.5\text{-}8\text{\,keV}}>3\times10^{42}\text{\,erg\,s}^{-1}$). The latter fraction is similar for star-forming and quiescent galaxies, and does not depend on galaxy stellar mass, suggesting that perhaps luminous AGNs are triggered by external effects such as mergers. We detect significant mean X-ray signals in stacked images for both the individually non-detected star-forming and quiescent galaxies, with spectra consistent with star formation only and/or a low-luminosity AGN in both cases. 
Comparing star formation rates inferred from the 2-10\,keV luminosities to those from rest-frame IR+UV emission, we find evidence for an X-ray excess indicative of low-luminosity AGNs. Among the quiescent galaxies, the excess suggests that as many as $70\%-100\%$ of these contain low- or high-luminosity AGNs, while the corresponding fraction is lower among star-forming galaxies ($43\%-65\%$). 
Our discovery of the ubiquity of AGNs in massive, quiescent $z\sim2$ galaxies provides observational support for the importance of AGNs in impeding star formation during galaxy evolution.
\end{abstract}


\section{Introduction}
Massive galaxies at $z\sim2$ are likely progenitors of today's massive ellipticals \citep[e.g.,][]{dokkum10}.
Combining broadband UV--MIR photometry with high-resolution imaging has shown that a relation akin to the Hubble sequence is in place at $z\sim2$, dividing massive galaxies into two main categories: star-forming galaxies with disk-like or irregular structures and quiescent galaxies with little or no star formation and extremely compact morphologies \citep[e.g.,][]{cimatti08,toft09,williams10,wuyts11,szomoru12}. The quiescent galaxies form a significant fraction ($30\%$--$50\%$) of all massive $\text{\it{z}}\sim2$ galaxies \citep[e.g.,][]{kriek06,williams09,toft09}, but how they got to this stage remains an open question.

Active galactic nuclei (AGNs) are one of the preferred hypothesized mechanisms for arresting star formation, as radiation, winds and jets from the central accreting black hole can remove the gas or heat it so that contraction cannot take place (see \citealt{fabian12} for a review of observational evidence). Semi-analytical simulations by, e.g., \cite{dimatteo05} and \cite{croton06}, show how reasonable descriptions of the AGN feedback on star formation in elliptical galaxies can result in galaxies which resemble those observed in terms of, e.g., their mass function. This could be the main mechanism leading the massive galaxies at $z\sim2$ from the blue star-forming cloud through the ``green valley'' and onto the red and dead sequence \citep{schawinski07,hopkins08}. An AGN would be capable of heating or expelling the gas residing in the galaxy as well as the gas that is believed to flow in via cold streams and otherwise provide the dominant source of star formation at $z\sim2$--$3$ \citep{dekel09}. To help test this picture for massive galaxies at $z\sim2$ we need better observational constraints on the incidence of AGN and the connection to star formation in these galaxies. 

A general concern when estimating the star formation rate (SFR) by fitting the broadband spectral energy distribution (SED) of a galaxy with stellar population synthesis models, is the relative dominance of AGN versus star formation. The dominant contributor to the mid-infrared (MIR) light is from UV-light reprocessed by dust, but the UV light can be emitted by either young stars or an AGN. At $z\sim2$ it is not clear, without the aid of high spatial and spectral resolution, what is causing the observed MIR emission from massive galaxies; is dust in the central regions being heated by AGN activity, is dust across a larger region of the galaxy being heated by star formation, or is a combination of the two scenarios taking place?

X-ray emission, on the other hand, does not suffer from strong dust obscuration. Observing in X-ray can thus lift the apparent degeneracy in interpreting the origin of the MIR light, and thereby help to give the true AGN fraction of a sample of galaxies. A galaxy dominated by AGN activity can be distinguished from a galaxy dominated by star formation by having a harder X-ray spectrum if the AGN is obscured by dust, or by simply having a very high X-ray luminosity, $L_{0.5\text{-}8\text{\,keV}}$. The most heavily obscured ``Compton-thick'' AGNs, with column densities $N_\text{H}>10^{24}\:\text{cm}^{-2}$, might be missed by X-ray selection, but can instead be identified via an excess of MIR emission over that expected from purely star-forming galaxies \cite[e.g.,][]{daddi07b,treister09}.

The shape of the MIR spectrum can also reveal AGNs, as the intense nuclear emission re-emitted by dust leads to a power-law spectrum in the MIR. For this purpose, color cuts have been devised using the {\it Spitzer} data \citep[e.g.,][]{stern05,donley12}. While this technique has the capability of detecting even Compton-thick AGNs, otherwise missed in X-ray, it has a low efficiency for low- to moderate-luminosity AGNs and its robustness has yet to be verified at $z\gtrsim2$ \citep{cardamone08}. 

In a galaxy where star formation is dominant, the total hard band X-ray luminosity, $L_{2\text{-}10\text{\,keV}}$, can be used to estimate the SFR \citep[e.g.,][]{grimm03,ranalli03,lehmer10}. An AGN will reveal itself if the X-ray SFR inferred this way is much larger than the SFR derived from other tracers such as H$\alpha$, UV, or MIR luminosity. Also, if star formation dominates the radiation output, the $L_x\rightarrow\,$SFR conversion will give an upper limit on the SFR as has been obtained for sub-mm galaxies \citep{laird10} and for massive, star-forming galaxies at $1.2<z<3.2$ \citep{kurczynski12}.

At redshifts around $2$, high-ionization lines in the rest-frame UV can be used as AGN indicators, but X-ray observations remain a more efficient way of identifying AGNs \citep{lamareille10}. Several studies of massive $z\sim2$ galaxies have therefore been made with the aim of uncovering AGN fractions, using the {\it Chandra} X-ray observatory. \cite{rubin04} performed a study of $40$ massive ($M_{\ast}=(1$-$5)\times 10^{11}M_{\odot}$) red ($J_s-K_s\geq2.3$) galaxies at $z\gtrsim 2$ by analyzing a $91$ ks {\it Chandra} exposure. Roughly $5\%$ of these were found to host an AGN with intrinsic $L_{2\text{-}10\,\text{keV}}>1.2\times 10^{43}\,\text{erg\,s}^{-1}$. Assuming that the stacked X-ray signal from the remaining X-ray undetected galaxies in X-ray comes from star formation alone, they derived a mean SFR broadly consistent with the typical mean SFRs estimated from SED fits. \cite{alexander11} analyzed the 4Ms {\it Chandra} observations of $222$ star-forming BzK galaxies ($M_{\ast}\sim10^{10}\text{-}10^{11}M_\odot$, \citealp{daddi07b}) in the Chandra Deep Field-South (CDF-S). $10\%$(23) showed X-ray emission consistent with AGN dominance, of which $5\%$(11) were found to contain heavily obscured AGNs, $4\%$(9) to have luminous, unobscured AGNs, and $3$ out of $27$ low-luminosity systems showed excess variability over that expected from star formation processes, indicating that at least some low-luminosity (rest-frame $L_{2\text{-}10\text{keV}}<10^{43}\text{\,erg\,s}^{-1}$) systems may contain AGNs.

The aim of this paper is to determine the AGN fraction in massive $z\sim2$ galaxies, and reveal any differences between galaxies characterized as quiescent or star forming. We address the matter by analyzing the X-ray emission from a mass-complete ($M_{\ast}>5\times10^{10}M_\odot$) sample of $1.5\leq z\leq 2.5$ galaxies residing in the CDF-S. The CDF-S, observed for $4\,$Ms, is currently the deepest X-ray view of the universe and so provides the best opportunity to study high-$z$ galaxies across a relatively large area ($464.5\,\text{arcmin}^2$).

Following a description of the sample selection (Sections~\ref{samp1}) and of the X-ray data and analysis (Section~\ref{x0}), we determine AGN fractions from the X-ray spectra and compare the X-ray inferred SFRs to UV+IR estimates (Section~\ref{aha}). Results are presented and discussed in Section~\ref{result}, and in Section~\ref{sum}, we sum up the conclusions. Throughout the paper we assume a flat cosmology with $\Omega_m=0.3$, $\Omega_\Lambda=0.7$ and $h=0.72$, and magnitudes are in the AB system.

\section{Galaxy sample}
\label{samp1}
We select our galaxies from the FIREWORKS\footnote{\url{http://www.strw.leidenuniv.nl/fireworks/}} catalog \citep{wuyts08}, which covers a field situated within the CDF-S and contains photometry in $17$ bands from the $U$ band to the MIR. For this study we extract a mass-complete sample of $M_{\ast}>5\times10^{10}M_\odot$ galaxies at $1.5\leq z\leq 2.5$ in a way similar to that of \cite{franx08} and \cite{toft09}. We will use spectroscopic redshifts when available \citep{vanzella08,xue11}, and photometric redshifts from the FIREWORKS catalog otherwise. In order to maximize the signal-to-noise (S/N) on results from X-ray stacking \citep{zheng12} and ensure a relatively homogeneous PSF across the {\it Chandra} field employed, we consider only galaxies that lie within $6\arcmin$ of the average {\it Chandra} aimpoint. 

We adopt galaxy stellar masses from the SED fitting results of \cite{franx08} and \cite{toft09}. In short, the SED fits were carried out with models by \cite{bruzual03}, choosing the best fit from adopting three different star formation histories (a single stellar population with no dust, an exponentially declining star formation history of timescale $300\,\text{Myr}$ with dust, and a constant star formation history with dust). In cases where the spectroscopic redshift differed by more than $0.1$ from the original FIREWORKS photometric redshift, we redid the SED fits in FAST\footnote{\url{http://astro.berkeley.edu/~mariska/FAST.html}} using an exponentially declining star formation history with a range of possible timescales from $10\,\text{Myr}$ to $\sim1\,\text{Gyr}$. As a quality parameter of the SED modeling, we demand an upper limit on the reduced $\chi^2_{\nu}$ of $10$ on the best-fit model. The SED fits provide SFR estimates, but we will be using SFRs derived from rest-frame UV+IR luminosities (see Section \ref{aha1}) as these include obscured star formation and are subject to less assumptions. From the observed photometry, rest-frame fluxes in $U$, $V$, $J$ band and at $2800\text{\,\AA}$ have been derived using InterRest\footnote{\url{http://www.strw.leidenuniv.nl/~ent/InterRest}} \citep{taylor09}. 

We divide the resulting sample of $123$ galaxies into quiescent and star-forming galaxies using the rest-frame $U$, $V$ and $J$ (falling roughly into the observed $J$, $K$ and IRAC 4.5$\,\mu\text{m}$ bands at $z\sim2$) colors. Dust-free but quiescent galaxies differ from dust-obscured starburst galaxies in that they obey the following criteria by \cite{williams09}: 
\begin{align}
U-V&>1.3  \\
V-J&<1.6 \\
(U-V)&>0.88\times(V-J)+0.49 \label{uvj2}
\end{align}
The fraction of quiescent galaxies identified within our sample using this method is $22\%\pm5\%(27/123)$. This is rather low compared to the $30$--$50\%$ found by \cite{toft09} for the same redshift and mass limit, but under the requirement that the sSFR (=SFR/$M_\ast$) from SED fitting is less than $0.03\,\text{Gyr}^{-1}$. If applying this criterion, we would arrive at a fraction of $42\%\pm7\%(51/123)$. A possible reason for the discrepancy between the two methods may be that we are using the $UVJ$ criterion in the limits of the redshift range in which it has so far been established. As the redshift approaches $2$, the quiescent population moves to bluer $U-V$ colors, possibly crossing the boundaries of Equation (\ref{uvj2}). However, we prefer the $UVJ$ selection technique in contrast to a cut in sSFR, because rest-frame colors are more robustly determined than star formation rates from SED fits.

\section{X-ray analysis and stacking method}
\label{x0}
The raw X-ray data from the CDF-S survey consist of 54 individual observations taken between 1999 October and 2010 July. We build our analysis on the work of \cite{xue11}, who combined the observations and reprojected all images. They did so in observed full band ($0.5$--$8\,$keV), soft band ($0.5$--$2\,$keV) and hard band ($2$--$8\,$keV), and the resulting images and exposure maps are publicly available.\footnote{\url{http://www2.astro.psu.edu/users/niel/cdfs/cdfs-chandra.html}}
\begin{figure}[t!]
\centering
\hspace{.0em}\raisebox{0.cm}{\includegraphics[width=0.95\columnwidth]{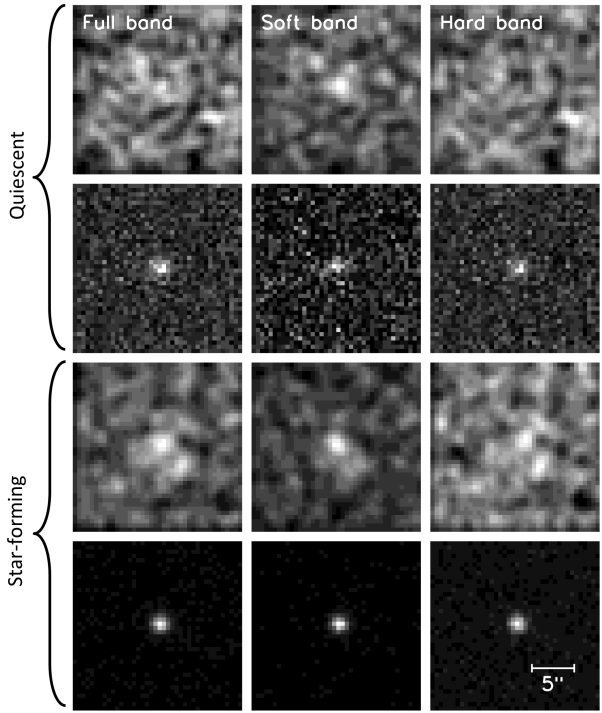}}
\caption{Stacked, background-subtracted and exposure-corrected $20\arcsec\times20\arcsec$ images in the three energy bands (columns) for the individually non-detected (top row) and detected (bottom row) quiescent and star-forming galaxies. Images of the non-detections have been smoothed with a Gaussian of width 2 pixels.}
\label{st_im}
\end{figure}
 \\
\noindent
We extract source and background count rates in the X-ray maps for all galaxies using the method of \cite{cowie12}. Source counts are determined within a circular aperture of fixed radii: $0\arcsec.75$ and $1\arcsec.25$ at off-axis angles of $\theta\leq3\arcmin$ and $\theta>3\arcmin$ respectively. Background counts are estimated within an annulus $8\arcsec$--$22\arcsec$ from the source, excluding nearby X-ray sources from the catalog of \cite{xue11}.

Each galaxy is classified as ``X-ray detected'' if detected in at least one band at $\geq 3\sigma$ significance, thereby creating  $4$ subsamples containing $8$ quiescent and detected, $19$ quiescent and undetected, $43$ star-forming and detected, and $53$ star-forming and undetected galaxies. 

While the X-ray detected galaxies can be analyzed individually, we stack the X-ray non-detected galaxies in order to constrain the typical X-ray flux from these. Stacked X-ray images of the $4$ subsamples are shown in Figure \ref{st_im} with all galaxies aligned to their $K_s$ band galaxy center positions from FIREWORKS. Representative count rates and associated errors for these stacks are calculated using the optimally weighted mean procedure of \cite{cowie12} and tabulated in Table \ref{counts_da} together with S/N values. 

\begin{table}[htbp]
\centering
\begin{tabular*}{\columnwidth}{p{0.2cm}p{2.0cm}p{2.0cm}p{2.0cm}} \toprule
		& \multicolumn{3}{c}{Count Rate $\pm1\sigma$ ($10^{-6}\,\text{s}^{-1}$) in } 	\\
 		& Full Band	 		& Soft Band 			& Hard Band 	\\ \hline
$Q$		& $0.71\pm0.96$  	&  $2.01\pm0.41$   	&  $-1.67\pm0.86$\\ 
		& $(0.7)$  				&  $(4.9)$   			&  $(-1.9)$\\ 
SF		& $2.51\pm0.55$  	&  $1.08\pm0.24$    	&  $1.29\pm0.49$\\
		& $(4.5)$  				&  $(4.5)$    			&  $(2.7)$\\ \hline
\end{tabular*}
\caption{\begin{flushleft}Stacking results for Quiescent ($Q$) and Star-forming (SF) galaxies not detected individually in X-Rays (in Parentheses the corresponding S/N)\end{flushleft}}
\label{counts_da}
\end{table}

The reliability of our chosen method for X-ray stacking and source count extraction was tested with Monte Carlo (MC) simulations. With $500$ MC simulations of $53$ (corresponding to the number of star-forming galaxies not detected in X-rays) randomly selected positions having no X-ray detections nearby, we obtain a histogram of stacked S/N values that is well fitted by a normal distribution with a center at $-0.02\pm0.8$, that is, consistent with no detection. Similar results were obtained using a sample size of $19$, the number of quiescent non-detections.

\begin{table*}[htbp]
\centering
\begin{tabular*}{15cm}{p{2cm}p{3cm}p{9cm}} \toprule 
$L_{0.5\text{--}8\text{\,keV}}$ ($\text{\,erg\,s}^{-1}$)		& 	HR					&	Classification \\ \hline
$>3\times10^{42}$			&	$<-0.2$					& Unobscured AGNs ($N_\text{H}<10^{22}\:\text{cm}^{-2}$)		\\
$>3\times10^{42}$			&	$>-0.2\text{ and }<0.8$		& Moderately obscured AGNs ($10^{22}<N_\text{H}<10^{24}\:\text{cm}^{-2}$)		\\
$>3\times10^{42}$			&	$>0.8$					& Compton-thick AGNs ($N_\text{H}>10^{24}\:\text{cm}^{-2}$)		\\
$<3\times10^{42}$ 			&	$<-0.2$					& Star-forming galaxy \\
$<3\times10^{42}$ 			&	$>-0.2$					& Low-luminosity obscured AGNs or star-forming galaxy\\ \hline
\end{tabular*}
\caption{\begin{flushleft}Classification scheme used in this work, with limits from \cite{szokoly04}, \cite{wang04}, and \cite{treister09}\end{flushleft}}
\label{class}
\end{table*}

We quantify the hardness of the X-ray spectra using the hardness ratio, HR$=(H-S)/(H+S)$,\footnote{Hardness ratio has the advantage, in comparison to $\Gamma$, of avoiding any assumptions regarding the shape of the spectrum.} where $H$ and $S$ are the net counts in the hard and soft band, respectively. Assuming a power-law spectrum, we also derive a photon index, $\Gamma,$\footnote{The photon index is defined as the exponent in the relation between photon flux density and energy: $dN/dE [\text{photons cm}^{-2}\text{ s}^{-1}\text{ keV}^{-1}]\propto E^{-\Gamma}$} with the online mission count rate simulator WebPIMMS\footnote{\url{http://heasarc.nasa.gov/Tools/w3pimms.html}} using a Galactic \ion{H}{1} column density of $8.8\times10^{19}\text{\,cm}^{-2}$ \citep{stark92} and not including intrinsic absorption. Whenever the S/N in either soft or hard band is below $2$, we use the corresponding $2\sigma$ upper limit on the count rate to calculate a limit on both HR and $\Gamma$, leading to a limit on the luminosity as well. When neither a hard- nor a soft-band flux is detected with an S/N above $2$, a typical faint source value of $\Gamma=1.4$ \citep{xue11} is assumed, corresponding to HR$\,\sim-0.3$ for an intrinsic powerlaw spectrum of $\Gamma=1.9$ \citep{wang04}.

We derive the unabsorbed flux from the count rate by again using the WebPIMMS tool, now with the $\Gamma$ tied to the observed value of either the individually detected galaxy or the stack in question. With this method, a count rate of $\sim 10^{-5}\text{\,counts\,s}^{-1}$ corresponds to a flux of nearly $10^{-16}\text{\,erg\,cm}^{-2}\text{\,s}^{-1}$ in full band at a typical value of $\Gamma=0.8$. Finally, the flux is converted into luminosity in the desired rest-frame bands using XSPEC version 12.0.7 \citep{arnaud96}.

\section{Interpretation of X-ray properties}
\subsection{Luminous AGN Identification} 
\label{aha}

\begin{figure}[htbp] 
\begin{center} $
\begin{array}{cc}
\includegraphics[width=0.84\columnwidth]{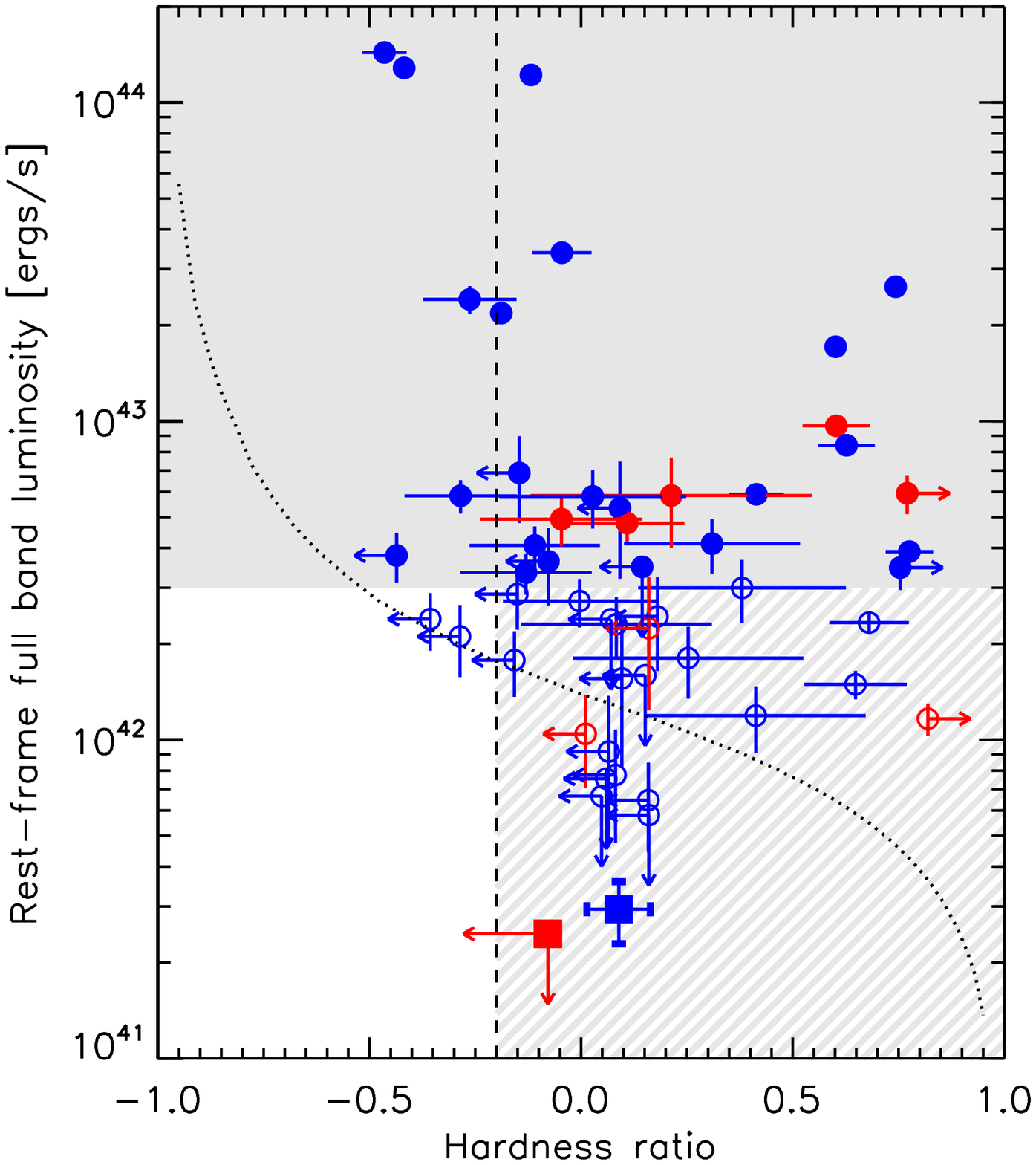} \\
\includegraphics[width=0.84\columnwidth]{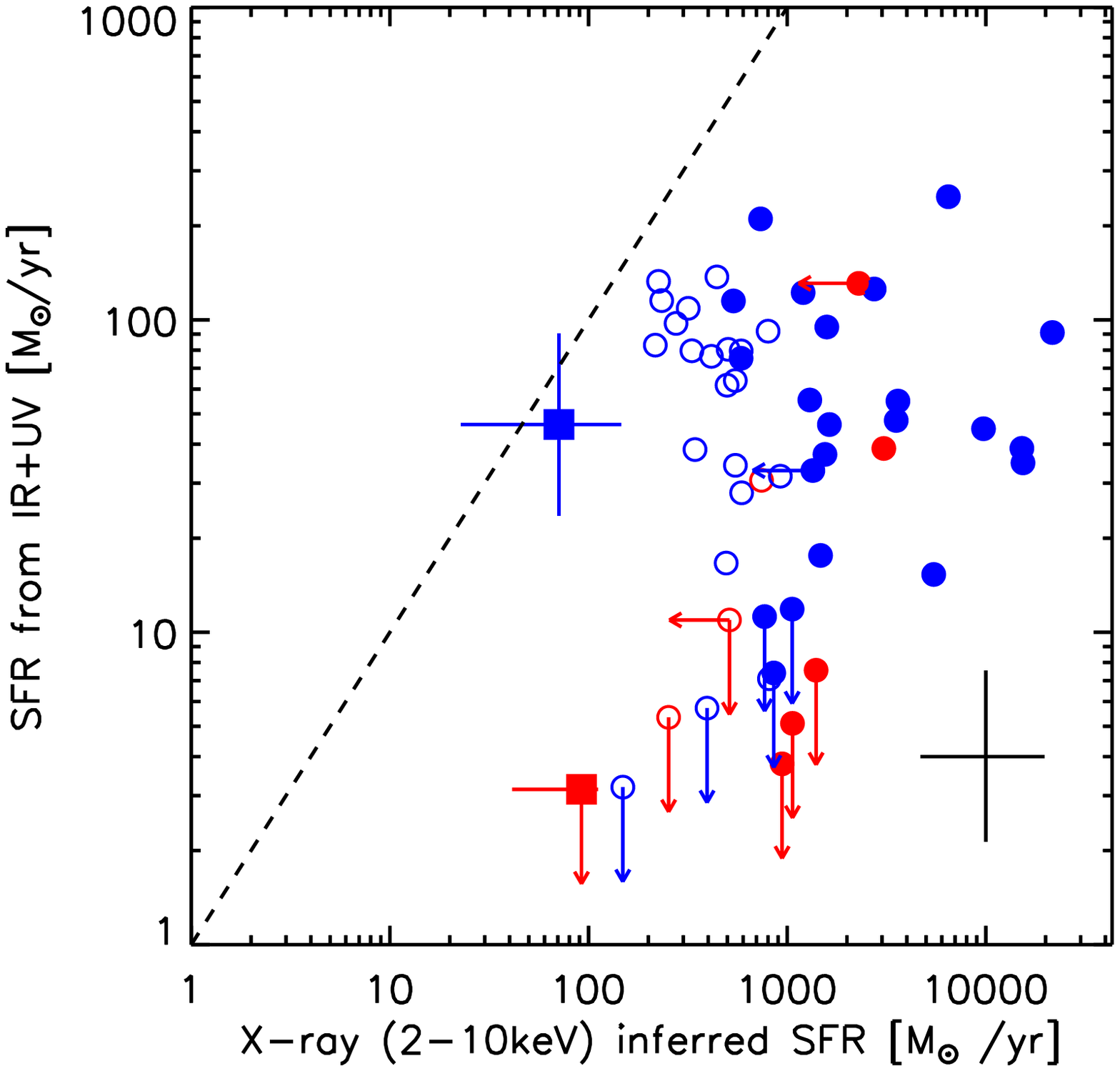}
\end{array} $
\end{center}
\caption{Top: $L_{0.5\text{--}8\,\text{keV}}$ vs. HR for all galaxies. Red: quiescent galaxies. Blue: star-forming galaxies. Squares: stacks of non-detected samples. Filled circles: luminous AGNs (selected with the X-ray criteria indicated by a shaded area). Open circles: galaxies dominated by low-luminosity AGNs or star formation processes alone (selected with X-ray criteria indicated by a hatched area). Dashed line: separating obscured ($N_\text{H}\gtrsim10^{22}\:\text{cm}^{-2}$) from unobscured AGNs. Dotted line: typical detection limit. Errorbars display 2$\sigma$ errors and limits are indicated with arrows. Bottom: SFR$_\text{UV+IR}$ vs. SFR$_{2\text{--}10\text{keV}}$ (see the text). Dashed line: equality. Typical errors on X-ray detected galaxies are shown in the lower right corner. Symbols are as above.}
\label{hr_lum}
\end{figure}

In the top panel of Figure \ref{hr_lum} are shown the rest-frame X-ray full band luminosities, $L_{0.5\text{--}8\,\text{keV}}$, not corrected for intrinsic absorption, versus observed hardness ratio, HR, for all X-ray detected galaxies as well as stacked non-detections. A typical detection limit on $L_{0.5\text{--}8\,\text{keV}}$ has been indicated with a dotted line in Figure \ref{hr_lum}, calculated as two times the background noise averaged over all source positions. A few detected galaxies are found below this limit due to these residing at relatively low $z$ and/or in regions of lower-than-average background. Galaxies with $L_{\text{0.5\text{--}8\,keV}}>3\times10^{42}\text{\,erg\,s}^{-1}$ are selected as luminous AGN, since star-forming galaxies are rarely found at these luminosities \citep{bauer04}. In addition, we adopt the HR criteria in Table \ref{class} in order to identify obscured and unobscured AGNs.

About $53\%(27/51)$ of the X-ray detected galaxies are identified as luminous AGNs, with $22$ moderately to heavily obscured ($-0.2<\;$HR$\;<0.8$) and $5$ unobscured (HR$\,<-0.2$) AGNs. The rest have X-ray emission consistent with either low-luminosity AGNs or star formation. We do not identify any Compton-thick AGNs directly, but three galaxies have lower limits on their hardness ratios just below HR$\;=0.8$, thus potentially consistent with Compton-thick emission (see Table \ref{class}). In total, we detect a luminous AGN fraction of $22\%\pm5\%(27/123)$, a fraction which is $23\%\pm5\%(22/96)$ for the star-forming and $19\%\pm9\%(5/27)$ for the quiescent galaxies.

Stacking the non-detections results in average X-ray source properties that exclude high luminosity AGNs. However, the inferred limits on their HR are consistent with a contribution from low-luminosity AGNs, the importance of which cannot be determined from this plot alone.

\subsection{X-ray Inferred SFR}
\label{aha1}

From the rest-frame hard band luminosity, $L_{2\text{--}10\,\text{keV}}$ it is possible to derive estimates of the SFR, as the number of high-mass X-ray binaries (HMXBs) is proportional to the SFR \citep{ranalli03}. \cite{kurczynski12} made a comparison of different SFR indicators, by applying three different $L_x\rightarrow\,$SFR conversions to $510$ star-forming BzK galaxies at $1.2<z<3.2$, selected as having $L_{\text{2\text{--}10\,keV}}<10^{43}\text{\,erg\,s}^{-1}$. While relations by \cite{persic04} and \cite{lehmer10} overestimated the true SFR (from rest-frame UV and IR light) by a factor of $\sim5$, the relation by \cite{ranalli03} provided a good agreement at $1.5<z<3.2$. But, as pointed out by \cite{kurczynski12}, all relations might lead to an overestimation due to contamination by obscured AGNs in the sample of SF galaxies. As we do not know the exact amount of obscured AGN contamination in the stacks, we have chosen to use the following relation by \cite{ranalli03} as a conservative estimate of the SFR: 
\begin{align}
\text{SFR}_{2\text{--}10\,\text{keV}}=2.0\times 10^{-40} \,L_{2\text{--}10\,\text{keV}}
\end{align}
with SFR measured in $M_\odot \text{\,yr}^{-1}$ and $L_{2\text{--}10\,\text{keV}}$ in $\text{erg\,s}^{-1}$. Another reason for not using the more recent relation by \cite{lehmer10} is that this relation was constructed for galaxies with SFR $>9\,M_{\odot}\text{\,yr}^{-1}$ only, whereas a large part of our sample has very low SFRs as inferred from SED fitting ($<1\,M_{\odot}\text{\,yr}^{-1}$). 
Following \cite{kurczynski12}, we use the observed soft band flux to probe the rest-frame hard band luminosity. The uncertainty on the SFR is estimated from the error on the observed soft-band flux together with a systematic error in the relation itself of $0.29\text{\,dex}$, as given by \cite{ranalli03}.

For comparison, the ``true'' SFR is inferred from rest-frame UV and IR light, SFR$_\text{UV+IR}$, following the method of \cite{papovich06}:
\begin{align}
\text{SFR}_{\text{UV+IR}}=1.8\times 10^{-10} (L_{\text{IR}}+3.3L_{2800})/L_\odot
\end{align}
where $L_{\text{IR}}$ is the total infrared luminosity and $L_{2800}$ is the monochromatic luminosity at rest-frame $2800\,\text{\AA}$. $L_{2800}$ comes from the rest-frame UV flux, $f_{2800\text{\AA}}$ (see Section \ref{samp1}), and in this context, the errors on $f_{2800\text{\AA}}$ are negligible. We derive $L_{\text{IR}}$ from the observed $24\,\mu\text{m}$ flux, $f_{24\mu\text{m}}$, using redshift-dependent multiplicative factors, $a(z)$, from the work of \cite{wuyts08}, Section 8.2 in that paper):
\begin{align}
L_{\text{IR}}[L_{\odot,8\text{--}1000\mu \text{m}}]=10^{a(z)}\cdot f_{24\mu\text{m}} 
\end{align}
The errorbars on $L_{\text{IR}}$ derive from the errors on $f_{24\mu\text{m}}$ and we further assume an uncertainty of $0.3$ dex in the relation of \cite{papovich06} as found by \cite{bell03} when comparing to H$\alpha$- and radio-derived SFRs.  The X-ray undetected quiescent galaxies are not detected with S/N$\,>3$ in $24\,\mu\text{m}$ either (except in one case) leading to a mean flux of $1.2\pm1.7\mu\text{Jy}$, of which we adopt a 2$\sigma$ upper limit in the following.

In the bottom panel of Figure \ref{hr_lum}, SFR$_{2\text{--}10\,\text{keV}}$ is compared to SFR$_\text{UV+IR}$ with the dashed line indicating equality. It is no surprise that nearly all of the individual X-ray detections have very high SFR$_{2\text{--}10\,\text{keV}}$ as compared to SFR$_\text{UV+IR}$, as we are largely insensitive to individual purely star-forming galaxies, given the detection limits in the top panel of Figure \ref{hr_lum} and the criteria in Table \ref{class}. 

For the star-forming stack, the X-ray inferred SFR of $71\pm51M_{\odot}\text{\,yr}^{-1}$ is consistent with the IR+UV inferred SFR of $46\pm33M_{\odot}\text{\,yr}^{-1}$, whereas the quiescent stack shows an SFR$_{2\text{--}10\,\text{keV}}$ of $92\pm65M_{\odot}\text{\,yr}^{-1}$ ($89\pm17M_{\odot}\text{\,yr}^{-1}$ when bootstrapping this sample $200$ times), well exceeding SFR$_\text{UV+IR}\leq3M_{\odot}\text{\,yr}^{-1}$.

\pagebreak
\section{Results and discussion}
\label{result}
\subsection{Luminous AGN Fraction}
In total, $22$ X-ray detected galaxies have emission consistent with containing a luminous (rest-frame $L_{0.5\text{--}8\text{keV}}>3\times10^{42}\text{\,erg\,s}^{-1}$) and obscured ($10^{22}<N_\text{H}<10^{24}\:\text{cm}^{-2}$) AGN, and a further five have emission consistent with a luminous unobscured ($N_\text{H}<10^{22}\:\text{cm}^{-2}$) AGN. This leads to a luminous AGN fraction of the full sample of $22\%pm5\%(27/123)$ and of the detected galaxies only, $53\%\pm13\%(27/51)$. The AGN fraction among both quiescent and star-forming galaxies, according to their X-ray spectra, is measured to be around $20\%$ as Table \ref{numbers} shows, meaning that AGNs in massive $z\sim2$ galaxies, even quiescent ones, are common, as proposed by \cite{kriek09} who studied the near-IR spectrum of one quiescent galaxy. 

\begin{table}[htbp]
\centering
\begin{tabular}{l|c|c} \toprule
							& 		Quiescent (27)			& Star-forming (96)			\\ \hline 
Luminous AGNs					&			$5$				& $22$		\\
\multirow{2}{*}{Low-luminosity AGNs}	&			$2$ (det)			& $19$ (det)			\\
							&			$12$--$19$ (non-det)	& $0$--$21$ (non-det)			\\ \hline
Luminous AGN fraction			&			$19\%\pm9\%$		& $23\%\pm5\%$				\\ 
Total AGN fraction				&			$70\%$--$100\%\,$		& $43\%$--$65\%\,$				\\ \hline
\end{tabular}
\caption{\begin{flushleft}X-Ray derived AGN Numbers and Fractions for Quiescent and Star-forming Galaxies, Divided into Luminous AGNs and Detected and Non-detected Low-luminosity AGNs\end{flushleft}}
\label{numbers}
\end{table}

This luminous AGN fraction is high when compared to the $5\%$ found by \cite{rubin04} (see the introduction), but this is likely a consequence of their 4$\sigma$ detection limit of $1.2\times 10^{43}\,\text{erg\,s}^{-1}$ in rest-frame $2$--$10\,\text{keV}$ being about twice as high as our limit of $5.5\times 10^{42}\,\text{erg\,s}^{-1}$ (at $\Gamma=1.4$), as calculated from the average background noise in the observed soft band. Adopting the detection limit of \cite{rubin04} and requiring an S/N of at least $4$, we reduce our fraction of luminous AGN to $8\%\pm3\%(10/123)$, consistent with the results of \cite{rubin04}. \cite{alexander11}, using also the $4\,$Ms CDF-S data, found a much lower X-ray detection fraction of $21\%\pm3\%$ as compared to ours ($53\%\pm7\%$), and a luminous AGN fraction of only $9\%\pm2\%(20/222)$. We believe that the discrepancies have several reasons, the main ones being: (1) our use of a mass-complete sample, whereas the BzK selection technique used by \cite{alexander11} includes galaxies down to $M_{\ast}\sim10^{10}M_{\odot}$ for which the total AGN fraction, assuming a fixed Eddington ratio, is expected to be lower above our detection limit,\footnote{For a fixed Eddington ratio, and assuming that galaxy bulge mass increases with total stellar mass, the AGN X-ray luminosity is expected to scale with galaxy stellar mass according to the $M_{\text{bh}}$--$M_{\text{bulge}}$ relation \citep{haring04}} (2) our updated source count extraction and stacking method leading to higher S/N, and (3) the use of $\Gamma$ instead of HR in the AGN identification conducted by \cite{alexander11}. For comparison, \cite{tanaka12} recently discovered a group of quiescent galaxies at $z=1.6$ with only one ``main-sequence'' star-forming galaxy. This group differed from local groups in having a remarkably high AGN fraction of $38^{+23}_{-20}\%$, consistent with our result, and which they interpret as possible evidence for AGN activity quenching star formation. 

\subsection{Importance of Low-luminosity AGNs}
As seen in Figure \ref{hr_lum}, the X-ray-identified luminous AGNs in general show an excess in SFR compared to that inferred from IR+UV emission. Among the galaxies classified as being dominated by either low-luminosity AGN or star formation, about $\sim90\%(21/24)$ have SFR$_{2\text{--}10\,\text{keV}}$ more than 1$\sigma$ above SFR$_\text{UV+IR}$. 

Surprisingly, the quiescent stack also has a much larger SFR$_{2\text{--}10\,\text{keV}}$ than SFR$_{\text{IR+UV}}$. Even when removing the marginally undetected galaxies with $2<$S/N$<3$, the resulting SFR$_{2\text{--}10\,\text{keV}}=62\pm19M_{\odot}\text{\,yr}^{-1}$ is still more than 3$\sigma$ above SFR$_{\text{IR+UV}}$. This discrepancy is only further aggravated if instead assuming the SFR$-L_{X}$ relation of \cite{lehmer10}. If caused by obscured star formation, we would have expected an average $24\,\mu\text{m}$ flux of $90\,\mu\text{Jy}$ for the individual galaxies, in order to match the lower limit on SFR$_{2\text{--}10\,\text{keV}}$. This is far above the upper limit of $3.4\,\mu\text{Jy}$ for the stack, suggesting that the X-ray flux of this stack is instead dominated by low-luminosity AGNs, but that their contribution to the $24\,\mu\text{m}$ flux remains undetectable in the {\it Spitzer}--MIPS data. 

We derive a lower limit on the low-luminosity AGN contribution for this stack of $19$ objects by constructing a mock sample of the same size and with the same redshifts, but containing {\sc n} low-luminosity AGNs with luminosities $L_{0.5\text{--}8\,\text{keV}}=10^{42}\text{\,erg\,s}^{-1}$ just below the detection limit and $19-\text{{\sc n}}$ galaxies with $L_{0.5\text{--}8\text{\,keV}}=10^{41}\text{\,erg\,s}^{-1}$ (all of them with $\Gamma=1.4$). From random realizations of this mock sample we find that at least $\text{{\sc n}}=12$ low-luminosity AGNs are required to match the observed rest-frame $2$--$10$ keV luminosity of the stack. Similarly, we find that by removing at least $12$ randomly selected galaxies it is possible to match the low SFR predicted by IR+UV emission, though only with a small probability ($\sim1\%$ of $200$ bootstrappings). In the following, we will adopt $63\%(=12/19)$ as a conservative lower limit on the low-luminosity AGN fraction among quiescent X-ray non-detections, but we note that the data are consistent with all the quiescent galaxies hosting low-luminosity AGNs if the luminosity of these is assumed to be only $L_{0.5\text{--}8\,\text{keV}}=7\times10^{41}\text{\,erg\,s}^{-1}$. 

Interestingly, the quiescent stack is only detected in the soft band, cf. Table \ref{counts_da}, whereas one might expect a significant contribution to the hard band flux from a low-luminosity AGN population as the one proposed above. The lack of a hard band detection can be explained by the fact that the sensitivity of the CDF-S observations drops about a factor of six from the soft to the hard band \citep{xue11}, meaning that the low-luminosity AGN population must be relatively unobscured ($\Gamma>1$, consistent with the value of $1.4$ assumed here), as it would otherwise have been detected in the hard band with an S/N$\,>2$.

The SFR$_{2\text{--}10\,\text{keV}}$ of the star-forming stack is consistent with its SFR$_\text{UV+IR}$, meaning that a strict lower limit to the low-luminosity AGN fraction here, is zero. However, performing the test above with the same model parameters, a maximum of $40\%(21/53)$ low-luminosity AGNs is possible, before the X-ray inferred SFR exceeds the upper limit on SFR$_\text{UV+IR}$.

It should be mentioned that the $24\,\mu\text{m}$ flux, especially at these high redshifts, is an uncertain estimator of the total rest-frame infrared luminosity, i.e., the entire dust peak, used in the conversion to SFR.  As shown by \cite{bell03}, one should ideally use the entire $8$--$1000\,\mu\text{m}$ range, e.g., by taking advantage of Herschel data, which can lead to systematic downward correction factors up to $\sim2.5$ for galaxies with  $L_\text{IR}\approx10^{11}L_{\odot}$ (similar to the inferred IR luminosities of our sample galaxies detected in $24\,\mu\text{m}$, showing a median of $L_\text{IR}=10^{11.5}L_{\odot}$) as demonstrated by \cite{elbaz10}. However, using the same conversion from $f_{24\mu\text{m}}$ to $L_\text{IR}$ as the one implemented in this study, \cite{wuyts11} showed that the resulting $L_\text{IR}$ for galaxies out to $z\sim3$ are consistent with those derived from PACS photometry with a scatter of $0.25$ dex. Hence, we do not expect the inclusion of Herschel photometry in the present study to significantly impact any of our results, and we leave any such analysis for future work.

Table \ref{numbers} gives an overview of the derived AGN fractions, both at high and low X-ray luminosity. Adding the numbers of luminous AGN, X-ray-detected low-luminosity AGNs as well as the estimated lower limit on the low-luminosity AGN fraction among non-detections, we arrive at a lower limit on the total AGN fraction of
\begin{align}
f_{\text{AGN}}\geq\frac{27+21+0.6\cdot19+0\cdot53}{123}=0.48
\end{align}
for all massive galaxies at $z\sim2$. While for the star-forming galaxies this fraction lies in the range from $43\%$--$65\%$, it must be $70\%$, and potentially $100\%$, for the quiescent galaxies. Using the upper limits on these numbers, a tentative upper limit on the total AGN fraction is $0.72$.\\

\subsection{Contribution from Hot Gas Halos}
We have so far considered star formation and AGN activity as causes of the X-ray emission observed, but a third possibility is an extended hot gas halo as seen around many nearby early-type galaxies in the same mass-range as our sample galaxies \citep{mulchaey10} and predicted/observed around similar spirals \citep{toft02,rasmussen09,anderson11,dai12}. AGN X-ray emission is expected to come from a very small, $R<1\,\text{pc}$, accretion disk surrounding the central black hole of the host galaxy \citep{lobanov07}, whereas very extended star formation in the galaxy or a hot gas halo surrounding it would lead to more extended emission. We investigate the possibilities for these latter cases by comparing radial surface brightness profiles of the $51$ individually X-ray detected galaxies out to a radius of $8\arcsec$ in both the stacked observed image and a correspondingly stacked PSF image. 

\begin{figure}[htbp] 
\centering
\includegraphics[width=\columnwidth]{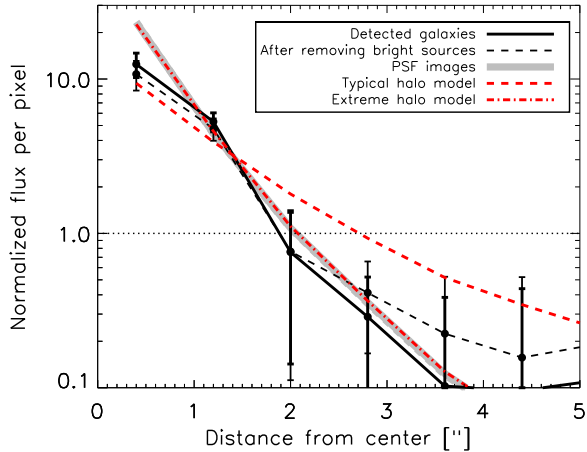}
\caption{Comparison of the radial surface brightness profiles in PSF images vs. observed images in full band for all detected galaxies. Black: stacked observed image centered on $K$-band positions from FIREWORKS for all detected galaxies (solid line) and if excluding the three X-ray brightest sources (dashed line). Grey line: stacked PSF image. Red: stack of halo model images, with $\beta=0.5$, $R_c=2.5\,\text{kpc}$ (dashed) and $\beta=1$, $R_c=1\,\text{kpc}$ (dot-dashed). }
\label{psf_im_is13}
\end{figure}

The profiles are calculated in full band only, because of the high S/N here as compared to the other bands (cf. Figure \ref{st_im}). For each galaxy, we extract the background subtracted source count per pixel within $10$ concentric rings of width $0\arcsec.8$ around the galaxy center positions from the FIREWORKS catalog, and each profile is normalized to the mean count rate in all rings. The same procedure is applied to the corresponding PSF images, extracted with the library and tools that come with CIAO,\footnote{\url{http://cxc.cfa.harvard.edu/ciao4.3/ahelp/mkpsf.html}} allowing for extraction at the exact source positions on the detector. We verified the robustness of using PSF models from the CIAO calibration database for this purpose, by repeating the above procedure for $51$ known X-ray bright point-like sources from the catalog of \cite{xue11}, and confirming that the resulting mean profile was fully consistent with that of the corresponding model PSFs.

As can be observed in Figure \ref{psf_im_is13}, the combined radial profile of our X-ray detected galaxies in the full band is consistent with the PSF of point sources stacked at the same detector locations. A Kolmogorov-Smirnov (K-S) test yields a statistic of $0.2$ and a probability of $99.96\%$ that the two profiles are drawn from the same parent sample.

Omitting the three sources with $L_{0.5\text{--}8\text{\,keV}}>10^{44}\text{\,erg\,s}^{-1}$ (cf. Figure \ref{hr_lum}) leads to a more extended profile (dashed black line in Figure \ref{psf_im_is13}), but the result still shows a high corresponding K-S probability of $70\%$. We also compare the profiles while recentering the images on the center of the X-ray emission as derived using WAVDETECT\footnote{WAVDETECT is part of the CIAO software.} instead of the $K$-band center positions listed in the FIREWORKS catalog, in order to test for the impact of any off-set between the X-ray and optical centroids. However, the recentering only resulted in small variations within the errorbars on the original profile. The same conclusions apply to the subsample of X-ray detected star-forming galaxies only, whereas the S/N was too low to perform a similar study on the quiescent, X-ray detected sample alone.

For comparison, we also simulated the stacked profile of extended hot gas halos, using for the surface brightness profile the $\beta$-model first introduced by \cite{cavaliere76}. With default parameters of $\beta=0.5$ and core radius $R_c=2.5\,\text{kpc}$, both taken from the study of $z\lesssim 0.05$ early-type galaxies by \cite{osullivan03}, the stacked profile of the halos, convolved with the corresponding PSFs, is well outside the errorbars of the measured profile. Only if assuming an extremely compact case of $\beta=1$ and $R_c=1\,\text{kpc}$, does the halo emission become sufficiently compact to mimic that of the PSF profile (see dash-dotted line in Figure \ref{psf_im_is13}), but the halo model then overpredicts the observed surface brightness on scales smaller than $1\arcsec$. In conclusion, a hot gas halo alone, described by a $\beta$-model, cannot explain the emission, unless one chooses model parameters that render the profile indistinguishable from a point-like source.

\subsection{Quenching of Star Formation by AGNs?}
It remains debated whether the presence of AGN is connected with internal galaxy properties associated with secular evolution or, to a higher degree, with external processes \citep{darg10}. For example, the cosmic star formation rate density and the number density of AGN share a common growth history with a peak in activity around $z\sim 2$--$3$ \citep{treister12a,dunlop11}, hinting at a co-evolution between SFR and supermassive black hole (SMBH) accretion \citep{schawinski11}. In this section, we will investigate the correlation, if any, between the presence of AGNs and internal/external processes, treating luminous and low-luminosity AGNs separately. As typical internal properties governing secular evolution we will focus on stellar mass, $M_\ast$, and SFR \citep{peng10}, while major mergers are taken as a likely case of external processes and as a phenomenon often associated with luminous AGNs \citep{treister12}. 

Starting with our X-ray identified luminous AGNs, we already found that the fraction of these does not correlate with star formation in our sample (cf. Table \ref{numbers}). In addition, the distribution of luminous AGNs and that of the rest with regards to their SFR$_{\text{IR+UV}}$ are similar, with a K-S test yielding a statistic of $0.21$ with a probability of $59\%$. Similarly, \cite{harrison12} found no correlation between AGN luminosity and quenching of star formation, albeit at a higher luminosity ($L_{2-8\,\text{keV}}>10^44 \text{erg\,s}^{-1}$) than probed here. Dividing the sample according to $M_\ast$ instead, by constructing two bins around the median mass of $1.1\times10^{11}\,M_\odot$, we arrive at similar luminous AGN fractions above and below this mass limit, namely $21\%\pm7\%(13/61)$ and $23\%\pm7\%(14/62)$, respectively. We thus find no clear evidence for the luminous AGN fraction of our sample to correlate with internal properties, suggesting that an external factor is of larger importance. 

An alternative is that luminous AGNs are primarily triggered by non-secular processes such as major mergers. \cite{treister12} found that major mergers are the only processes capable of triggering luminous ($L_{\text{bol}}\gtrsim10^{45}\,\text{erg\,s}^{-1}$) AGN at $0<z<3$. Our luminous AGN fraction is consistent with the major merger fraction of massive ($M_{\ast}>10^{11}M_{\odot}$) galaxies at $1.7<z<3$ found by \cite{man12} to be $15\%\pm8\%$ (compared to a luminous AGN fraction of $11\pm3\%(13/123)$ when excluding galaxies with $M_{\ast}<10^{11}M_{\odot}$). This is consistent with the idea that our luminous AGNs are triggered by major mergers, but a more direct test of this scenario would be to search for major mergers using the imaging data available in CDF-S. Indeed, \cite{newman12} used {\it HST} CANDELS imaging in the GOODS-South field residing within CDF-S (along with the UKIRT Ultra Deep Survey) to arrive at roughly equal pair fractions when comparing massive ($M_{\ast}>10^{10.7}M_{\odot}$) quiescent to massive star-forming galaxies at $0.4<z<2$. Again, this is in agreement with our result that the luminous AGN fraction does not vary between quiescent and star-forming galaxies. A detailed morphological study of our X-ray identified luminous AGN using the {\it HST}/WFC3 data to search for evidence of merging is an interesting extension to the work presented here, which we leave for a future study. 

Turning toward our low-luminosity AGNs, we {\it do} see X-ray evidence for an enhanced population among the quiescent galaxies when compared to their star-forming equivalents. The mean masses of our non-detected samples of quiescent and star-forming galaxies are similar ($14.6\pm1.6$ and $12.7\pm1.0\times10^{10}\,M_\odot$ respectively), suggesting that the relevant factor here is SFR. At $0.01<z<0.07$, \cite{schawinski09} found that massive ($M_{\ast}\gtrsim10^{10}M_{\odot}$) host galaxies of low-luminosity AGNs all lie in the green valley, that is, at some intermediate state after star formation quenching has taken place and before the SED is truly dominated by old stellar populations. The observation that these host galaxies had been quiescent for $\sim100\text{\,Myr}$, ruled out the possibility that one short-lived, luminous AGN suppressed star formation and, at the same time, made it unlikely that the same AGN quenching star formation was still active, given current AGN lifetime estimates of $\sim10^7$--$10^8$\text{\,yr} \citep{dimatteo05}. Rather, the authors favored a scenario in which a low-luminosity AGN already shut down star formation, followed by a rise in luminosity, making the AGN detectable. At $z\sim 2$, SED fits of massive ($M_{\ast}>10^{11}M_{\odot}$) quiescent galaxies show that the quenching typically took place $\sim1\;$Gyr before the time of observation (\citealp{toft12}; Krogager, J.-K. et al., in preparation), demanding an even longer delay {\it or} an episodic AGN activity as frequently applied in models \citep{croton06}. Episodic AGN activity could explain why we see evidence for a higher low-luminosity AGN fraction among quiescent as compared to star-forming galaxies, but it would also require the low-luminosity AGN phase in quiescent galaxies to last at least as long as the dormant phase. Future modeling and observations will show whether this is in fact possible. 

We conclude that our data are consistent with a scenario in which luminous AGNs in massive galaxies at $z\sim2$ are connected with major mergers or other non-secular processes, while the presence of low-luminosity AGNs in the majority of quiescent galaxies suggests that these AGNs present an important mechanism for quenching star formation and keeping it at a low level. Ultimately what happens at $z\sim2$ has to agree with the subsequent evolution that changes size and morphology of quiescent galaxies (Barro et al. 2012)

\section{Summary}
\label{sum}
Our main conclusions are on the following two topics: 
\begin{enumerate}
\item {\it Luminous AGN fraction}\\
We find a luminous AGN fraction of $22\%\pm5\%$ among massive ($M_{\ast}>5\times10^{10}M_\odot$) galaxies at redshifts $1.5 \leq z \leq 2.5$, using their X-ray properties extracted from the 4 Ms Chandra Deep Field South observations. Among the X-ray detected galaxies, $53\%\pm13\%$ harbor high-luminosity AGNs, while stacking the galaxies not detected in X-ray, leads to mean detections consistent with low-luminosity AGNs or pure star formation processes. The luminous AGN fraction among quiescent and star-forming galaxies is similar ($19\%\pm9\%$ and $23\%\pm5\%$, respectively) and does not depend on galaxy $M_{\ast}$. \\
We confirmed that extended X-ray emission from a hot gaseous halo is not a viable explanation for the observed X-ray emission of the X-ray detected galaxies.
\item {\it Limits on total AGN fraction}\\
We convert the rest-frame hard band X-ray luminosity into an upper limit on the star formation rate, SFR$_{2\text{--}10\,\text{keV}}$ and compare to that derived from the rest-frame IR+UV emission, SFR$_{\text{IR+UV}}$. All luminous AGNs show an excess in SFR$_{2\text{--}10\,\text{keV}}$ as expected, and so does a large fraction ($\sim90\%$) of the remaining detected galaxies. While the star-forming galaxies not detected in X-ray have a mean X-ray inferred SFR of $71\pm51M_{\odot}\text{\,yr}^{-1}$, consistent with their SFR$_{\text{IR+UV}}$, the stack of quiescent galaxies shows an excess in SFR$_{2\text{--}10\,\text{keV}}$ of a factor $>10$ above the upper limit on SFR$_{\text{IR+UV}}$. For these galaxies, we find that a minimum fraction of $\sim60\%$ must contain low-luminosity ($L_{0.5\text{--}8\,\text{keV}}\approx10^{42}\text{\,erg\,s}^{-1}$) AGNs if the SFR estimates from X-ray are to be explained, and that low-luminosity AGNs might be present in all of them. On the other hand, for the star-forming stack, we derive a low-luminosity AGN fraction of $0$--$40\%$. \\
Gathering all low- and high-luminosity AGNs, we derive a lower limit to the total AGN fraction of $48\%$, with a tentative upper limit of $72\%$.
\end{enumerate}


Our study is the first to present observational evidence that, at $z\sim2$, the majority of quiescent galaxies host a low- to a high-luminosity AGN, while the AGN fraction is significantly lower in star-forming galaxies. These findings are consistent with an evolutionary scenario in which low-luminosity AGN quench star formation via the energetic output from SMBH accretion, which, if believed to continue in an episodic fashion as often invoked by models, would need to have ``dormant'' phases at least as long as ``active'' phases.

We find that the high-luminosity AGNs are likely related to non-secular processes such as major mergers. In the future, examining the X-ray properties of galaxies in a larger sample, cross-correlated with signs of major mergers, may shed further light on the co-evolution of AGNs and host galaxy. 

\acknowledgments
\section*{Acknowledgments}
We thank the anonymous referee for careful reading of the manuscript and suggestions that improved the paper. We also enjoyed useful discussions with Ezequiel Treister, Marianne Vestergaard, Tomotsugu Goto and Harald Ebeling. We thank Stijn Wuyts and Natascha M. F\"{o}rster Schreiber for photometric redshifts and SED modeling results. We gratefully acknowledge the support from the Lundbeck foundation. J.R. acknowledges support provided by the Carlsberg Foundation. The Dark Cosmology Centre is funded by the Danish National Research Foundation.

\bibliographystyle{apj}
\bibliography{minbib}

\end{document}